\documentclass[%
 reprint,
 amsmath,amssymb,
 aps,
]{revtex4-2}

\usepackage{graphicx}
\usepackage{dcolumn}
\usepackage{bm}
\usepackage{braket}
\usepackage{amsmath}
\usepackage{subcaption}
\usepackage{tikz}
\usepackage[utf8]{inputenc}
\usepackage[T1]{fontenc}

\begin{document}

\preprint{APS/123-QED}

\title{Steganographic Entanglement Sharing}

\author{Bruno Avritzer}
\email{avritzer@usc.edu}
 \affiliation{Department of Physics and Astronomy, University of Southern California, Los Angeles, California}
\author{Todd A. Brun}%
 \email{tbrun@usc.edu}
\affiliation{%
 Ming Hsieh Department of Electrical and Computer Engineering, University of Southern California, Los Angeles, California
}%
\affiliation{Department of Physics and Astronomy, University of Southern California, Los Angeles, California}
\date{\today}

\begin{abstract}
    In a previous work \cite{priorwork} we have discussed a theoretical grounding for classical steganography using quantum Fock and coherent states in an optical channel, building on previous work by Wu et al \cite{Wu1,Wu2}. In that work, we discussed protocols which disguise  communications to mimic the thermal state of a harmonic oscillator. In this work we will extend this to transmission of quantum information, and demonstrate the utility of steganographic entanglement sharing in practical contexts like nonclassical state teleportation, even with the presence of an active eavesdropper.    
\end{abstract}
\maketitle


\section{\label{sec:intro}Introduction}
Steganography, meaning ``concealed writing'', refers to the practice of concealing information within an apparently harmless medium. For example, one might write a letter in which every third word can be combined to spell a hidden sentence. In this way, we can present a paragraph which is not suspicious to an eavesdropper intercepting our letter while still communicating our true message to another party with whom we have agreed a scheme in advance.  

``Concealed writing'' need not refer to literal writing, however. Steganography can be performed in any medium which can be used for communication. In \cite{priorwork} we considered a one-way optical channel through which various states of light, such as coherent states, could be transmitted, and in which a potential eavesdropper might expect to see only thermal noise. We showed in that work that the thermal state can be mimicked perfectly using either coherent or Fock states with an appropriate encoding --- this extends to any state which can be neatly diagonalized in the basis of coherent or Fock states, respectively. 

However, the individually generated Fock and coherent states discussed in the previous paper are only suitable for sending classical information. In this work, we will develop protocols for transmitting quantum information while mimicking a thermal state in the presence of an active eavesdropper. We will further analyze these protocols to determine under what monitoring conditions a quantum advantage is achievable, and quantify that advantage where possible. 

The protocols we will develop are not necessarily limited to the mimicry of thermal states. In principle, any single-mode state which is not a pure state can be mimicked by a subsystem of an entangled with a fidelity of 1state, and a channel which contains only pure states is not a physically realizable system as there will always be some noise. So it is always possible to share some entanglement covertly via steganography, even if the noise model is not thermal; the thermal state is just the subsystem which corresponds to the greatest amount of entanglement in the bipartite state, and also has the property of being easy to mimic with lab-generated states. We will focus our attention on the thermal noise model for the bulk of this paper for those reasons. 
\section{\label{sec:idle}Idle Channel}
A thermal state for a harmonic oscillator is given by
\begin{equation}
    \rho_{th} = \frac{1}{Z} \sum_{n=0}^\infty e^{-\frac{\hbar\omega (n+1/2)}{k_BT}}\ket{n}\bra{n} ,
\end{equation}
where 
\begin{equation}
    Z=\sum_{n=0}^\infty e^{-\frac{\hbar\omega(n+1/2)}{k_BT}}=\frac{1}{2}\text{csch}\left(\frac{\hbar\omega}{2k_BT}\right)
\end{equation} is the partition function. 

If we describe the thermal state of a mode in a channel in terms of the average number of photons transmitted, known as
\begin{equation}
    \Bar{n}=\left(e^{\frac{\hbar\omega}{k_BT}}-1\right)^{-1} ,
\end{equation}
we can reformulate the expression for $\rho_{th}$ in a simpler form:
\begin{equation}
    \rho_{th}=\frac{1}{\Bar{n}+1}\sum_{n=0}^\infty \left(\frac{\Bar{n}}{\Bar{n}+1}\right)^n\ket{n}\bra{n} .
\end{equation}
In this case, the thermal state is expressed in the Fock basis of harmonic oscillator energy eigenstates. The two mode squeezed vacuum (TMSV) state is a fidelity entangled state given by
\begin{equation}
    \rho_{\text{TMSV}} = \text{sech}^2(r)\sum_{m=0}^\infty\sum_{n=0}^\infty (\tanh(r))^{n+m}\ket{nn}\bra{mm}
\end{equation}
in the Fock basis $\ket{nn}=\ket{n}_A\ket{n}_B$, where $r$ is the squeezing parameter. It is maximally entangled in the sense that it is impossible to increase the entanglement level of the state using ``passive'' Gaussian operations, meaning those which preserve the photon number of the state. If we take the partial trace of the above over the first subsystem, we have
\begin{equation}
    \text{Tr}_A(\rho_{\text{TMSV}}) = \frac{1}{\cosh^2(r)}\sum_{n=0}^\infty \frac{\sinh^{2n}(r)}{\cosh^{2n}(r)}\ket{n}\bra{n}
\end{equation}
which is exactly a thermal state with $\Bar{n}=\sinh^2(r)$. As such, if we transmit one mode of the entangled state through a channel, it is identical to the thermal state with appropriate parameter and therefore a suitable primitive for steganographic entanglement sharing with a fidelity of 1.

There is a clear limitation, however: the $\Bar{n}$ of the channel we are trying to imitate sets a limit on how squeezed the resource state can be, and therefore on how much entanglement can be transmitted in each channel use. Specifically, for the entanglement of formation $E_F$ of a state we would like to mimic the thermal state, we have
\begin{equation}
    E_F\leq S(\rho_{th})=(\Bar{n}+1)\log(\Bar{n}+1)-\Bar{n}\log\Bar{n}
\end{equation}
where equality occurs for the TMSV state, which is pure. Likewise, if the channel state to be mimicked is a mixed state with entropy $S$ but not thermal, the above expression $E_F\leq S$ holds.   

\begin{figure}
    \centering
    \includegraphics[scale=.6]{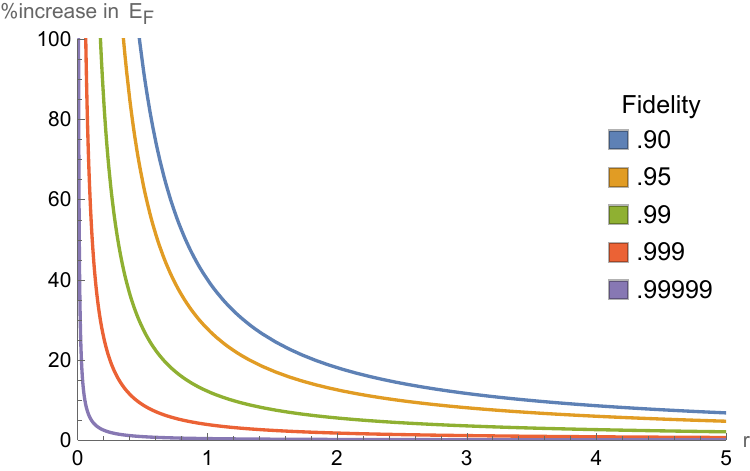}
    \caption{The percentage increase in the entanglement of formation as a function of the base squeezing level $r$ of the resource state. Note that $r$ rather than $\Bar{n}=\sinh^2(r)$ is the limiting parameter in the ``low-energy'' regime, as an $\Bar{n}$ of 20 corresponds to a squeezing level of $r\approx 2$. }
    \label{fig:entmarg}
\end{figure}

A natural concern is that the TMSV state may be prepared imperfectly due to experimental limitations, for example using a scheme such as that in \cite{sqexp}. The quality factor of the microring resonator may be degraded due to fabrication imperfections, so that the squeezing level is $r'$ rather than $r$. What then is the fidelity between the idealized thermal state and the actual thermal state produced? The answer is quite straightforward assuming only this limitation: 
\begin{equation}
    \sqrt{F}=\frac{1}{\sqrt{(1+\Bar{n})(1+\Bar{m})}-\sqrt{\Bar{n}\Bar{m}}}=\text{sech}(r-r')
\end{equation}
which is bounded below by $1-\frac{(r-r')^2}{2}$. This bound has another interpretation: it tells us how much fidelity we need to sacrifice to transmit a state which is more entangled than what the channel permits. For example, we can exceed the base squeezing level by up to $0.87\text{dB}$ of squeezing and still maintain a fidelity of .99. This increase in $E_F$ can be very significant, as can be seen in Figure \ref{fig:entmarg}. Additionally, since the fidelity is nondecreasing under partial trace (and the trace distance is likewise nonincreasing), the fidelity of TMSV state preparation provides another bound on the thermal state fidelity:
\begin{equation}
    F_{ch}\geq F_{\text{prep}},
\end{equation}
meaning that if it is possible to prepare a TMSV state with fidelity $F_{\text{prep}}$, the fidelity $F_{ch}$ of the channel state with the thermal state Eve is expecting is at least $F_{\text{prep}}$.  
\section{\label{sec:eve}Channel with Eavesdropper}
The above calculation would, in the classical case, constitute a suitable mapping for steganographic communication. However, since the resource we are sharing is an entangled state, there is an additional complication. Steganography assumes the existence of an eavesdropper, who can observe the channel in some way. This observation can have the effect of disturbing the transmission of entanglement, as often happens in a quantum key distribution protocol. Therefore, in this case it is more critical to provide some models of the eavesdropper and describe the efficacy of communication under those circumstances.
\subsection{\label{sec:werner}Werner Model of the Eavesdropper}
\begin{figure*}
\centering
\begin{subfigure}[b]{0.3\textwidth}
    \includegraphics[width=1.1\textwidth]{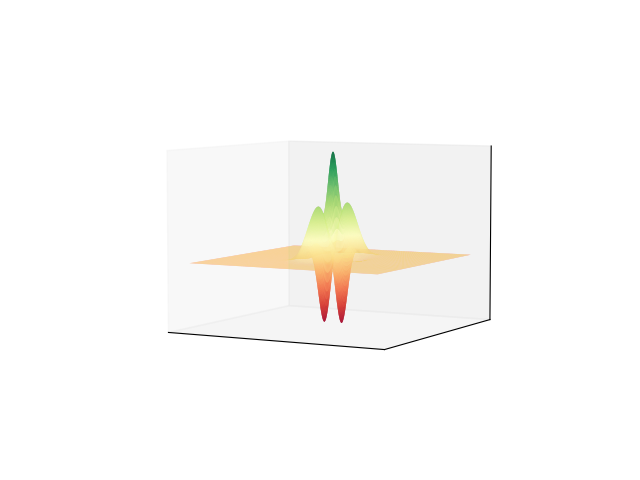}
    \caption{}
    \label{ref:cat1.5j}
\end{subfigure}
\hfill
\begin{subfigure}[b]{0.3\textwidth}
    \includegraphics[width=1.1\textwidth]{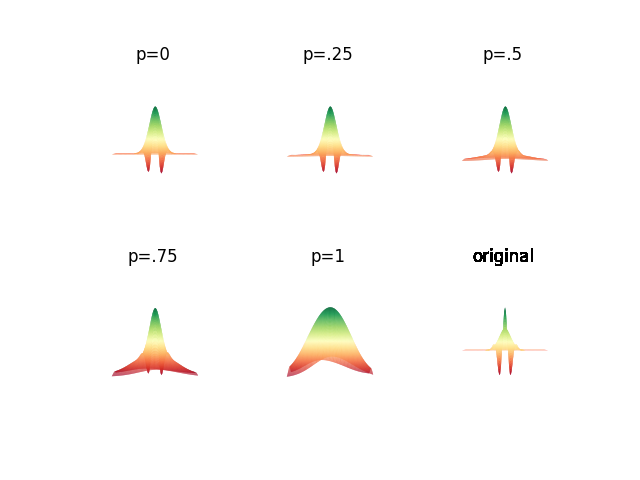}
    \caption{}
    \label{ref:wignerscattel}
\end{subfigure}
\hfill
\begin{subfigure}[b]{0.3\textwidth}
     \includegraphics[width=1.1\textwidth]{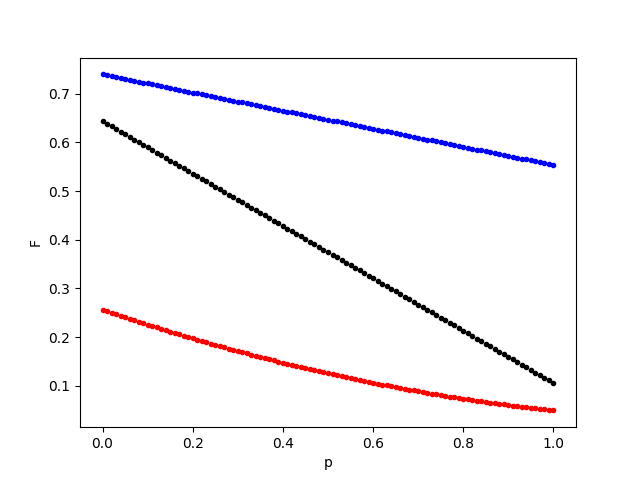}
     \caption{}
     \label{ref:fvdg}
\end{subfigure}
\hfill
   \caption{(a) the Wigner function of the odd cat state with $\alpha=-1.5i$, (b) average Wigner function and teleportation fidelity of the cat state with $\alpha=-1.5i$ using a TMSV state with squeezing parameter $r=1.15$ and noise channel $\xi_p$, and (c) the upper and lower Fuchs--Van de Graaf bounds \cite{FVDGbound} for the result of the teleportation using $\xi_p$. These results were obtained using the StrawberryFields package \cite{sfields1,sfields2}. A derivation of the Fuchs--Van de Graaf lower bound is provided in Appendix \ref{appendixa} (the upper bound does not hold formally except when $p=0$ but is provided for illustrative purposes). These bounds become tighter as the state becomes more and less pure, respectively.}
    
    \label{figure: cattel}
\end{figure*}

In the classical case we considered as an adversary an all-powerful eavesdropper, whose powers were restricted only by the laws of physics. This applies only to the eavesdropper's ability to perform arbitrary measurements on the channel state, of course. A comparably aggressive model is the probabilistic eavesdropper channel
\begin{equation}
    \xi_p(\rho_{AB}) = p\text{Tr}_B(\rho_{AB})\otimes \text{Tr}_A(\rho_{AB})+(1-p)\rho_{AB}
\end{equation}
which represents the full destruction of entanglement in the shared state when observed by the eavesdropper, an event that occurs with probability $p$. One interpretation of this is that Eve removes the state from the channel, perfectly determines (through unspecified means, as this is unphysical) what state she has acquired, and sends that state to Bob. Note that from Bob's point of view, it is impossible to determine whether Eve has tampered with the state at all. This is in contrast to a channel which sends the state to $\text{Tr}_B(\rho_{AB})\otimes\ket{0}\bra{0}$ rather than the thermal state: it is not only possible to detect an eavesdropper with this vacuum channel, but the vacuum state is often more useful as a resource for e.g. teleportation. Surprisingly, even with this aggressive channel active, it is possible to perform useful quantum tasks with the distributed entanglement even for relatively high values of $p$.   
\subsubsection{\label{sec:wtel}State Teleportation}
\begin{figure*}
    \centering
    \begin{subfigure}[b]{0.6\textwidth}
    \includegraphics[scale=.3]{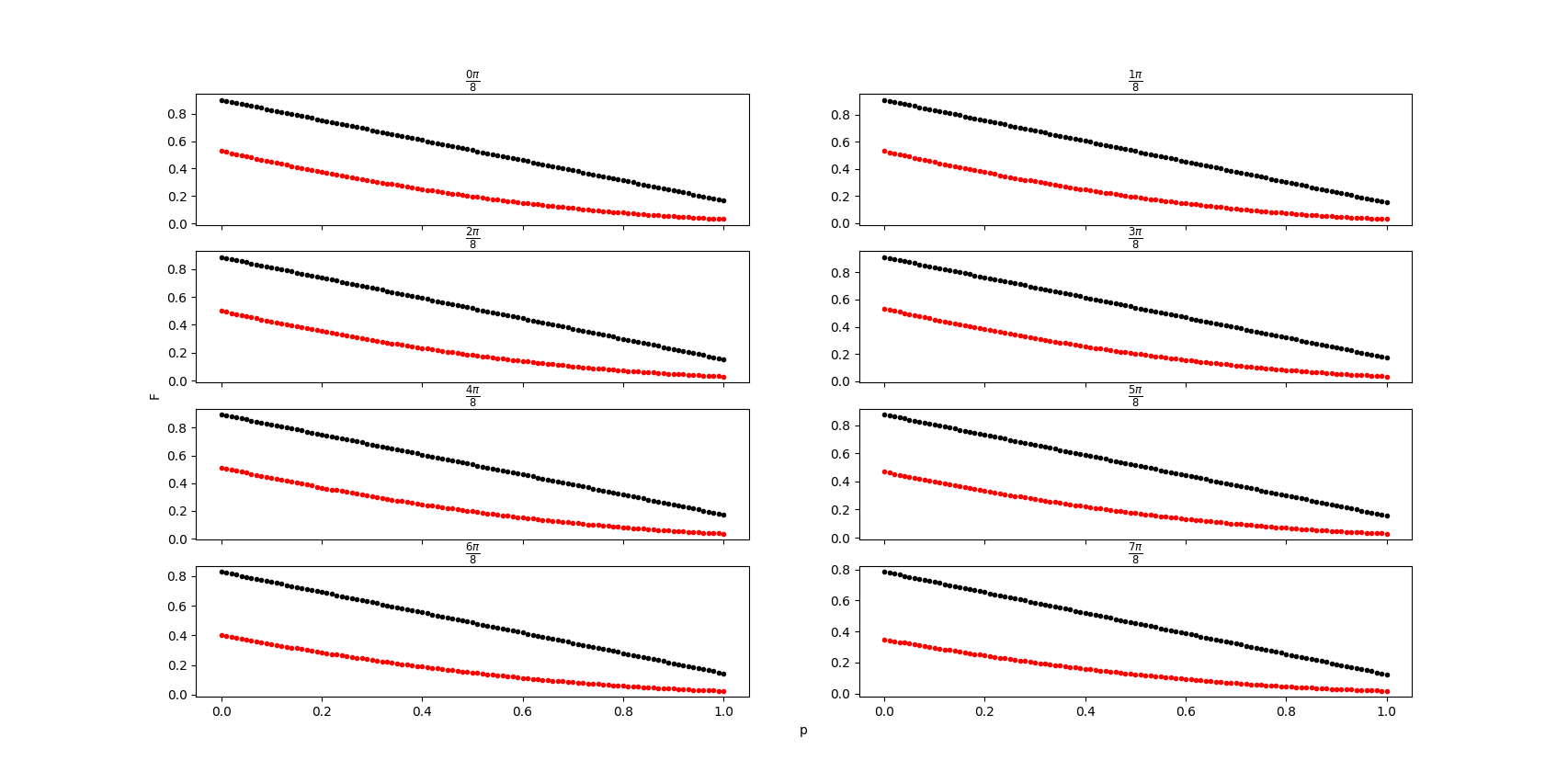}
    \caption{}
    \label{tel:gkpfid}
    \end{subfigure}
    \hfill
    \begin{subfigure}[b]{0.3\textwidth}
    \includegraphics[scale=.4]{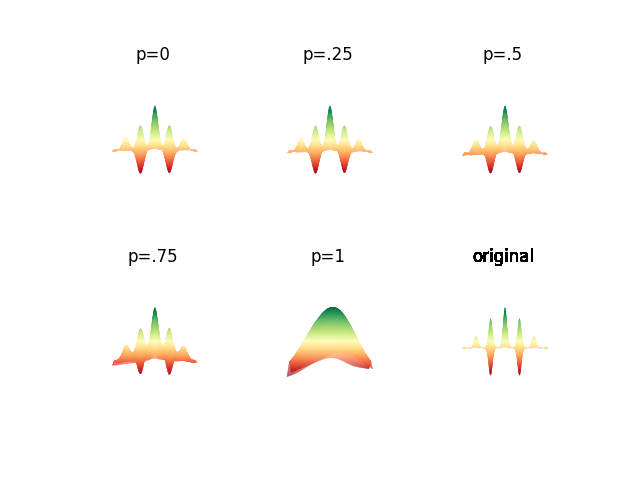}
    \caption{}
    \label{tel:gkpfid}
    \end{subfigure}
    \hfill
    \caption{(a) the fidelities of GKP state teleportation and (b) Wigner function for $\theta=0$  with $r=1.15$ using the channel $\xi_p$ for $\theta=\frac{n\pi}{8}$, where $\ket{\psi}=\cos\frac{\theta}{2}\ket{0}+\sin\frac{\theta}{2}\ket{1}$ (black) and the Fuchs-Van de Graaf lower bound (red). We can see that the average  teleportation fidelity is higher for the GKP state than for the above cat states at the same level of squeezing.}
\end{figure*}

In the case of TMSV state transmission, the channel above is explicitly given by
\begin{equation}
    \xi_p(\rho_{\text{TMSV}}) = p\rho_{th}\otimes\rho_{th}+(1-p)\rho_{\text{TMSV}},
\end{equation} which is a sort of continuous variable analog of the Werner channel \cite{Werner}, and has been studied previously in \cite{FWerner}. In that work, it was demonstrated that coherent state teleportation could be performed with an asymptotic (in $r$) average fidelity of $1-p$, which for $p<.5$ is an improvement over the classical case. It is remarkable that a channel which is so frequently and completely disturbed is still able to provide a quantum advantage!

It turns out that this result extends to the teleportation of nonclassical states. The (odd) cat state is a highly nonclassical state which is given by
\begin{equation}
    \ket{\psi(\alpha)}=\frac{\ket{\alpha}-\ket{-\alpha}}{N}
\end{equation} where $\ket{\alpha}$ is a coherent state and $N$ is the appropriate normalization. Such a state can be used for bosonic error correction \cite{eccat} and for Gottesman-Kitaev-Preskill (GKP) state preparation \cite{catgkp}. This continuous variable state can be teleported using the Braunstein-Kimble teleportation scheme \cite{BKTel}, which is analogous to the qubit teleportation scheme. First, a maximally entangled TMSV state is prepared. Then, one mode of that state is interfered with the state to be teleported using a 50-50 beamsplitter. That state is projected onto a maximally entangled basis, in this case by measuring the $x$ and $p$ quadratures of the two output ports. Finally, the measurement results are used to apply a correction to the state of the other mode using a displacement operator. Although the steps are similar to qubit teleportation, there is a flaw: the TMSV state is not maximally entangled in the $x$-$p$ quadrature basis except in the limit of infinite squeezing (it is maximally entangled only for a given level of squeezing). There is a second flaw: projections onto the Bell basis produce a discrete outcome, while quadrature measurements produce a continuous outcome which is less robust to error. These factors make continuous variable teleportation somewhat less reliable than discrete variable teleportation with comparable resources.

A key signature of the cat state's nonclassical nature is the negativity of the cat state Wigner function. Figure \ref{figure: cattel} shows that for values of $p$ up to $.5$ the Wigner function shows robust negativity after being teleported, which is not achievable by classical means. This indicates that the average performance of this highly noisy channel is significantly better for this task than an unmonitored classical channel, up to about $p=.5$. The teleportation fidelity in this case is given by 
\begin{equation}
    \begin{aligned}
        F=pF_{th}+(1-p)F_{\text{TMSV}}
    \end{aligned}
\end{equation}
where 
\begin{equation}
    F_{\text{TMSV}}=\frac{1}{1+e^{-2r}}-\frac{1+e^{-4z^2}-e^{\frac{-4e^{-2r}z^2}{1+e^{-2r}}}-e^{\frac{-4z^2}{1+e^{-2r}}}}{2(1+e^{-2r})(1-e^{-2z^2})^2}
\end{equation} is the fidelity of teleportation using the TMSV state \cite{BKTel} and 
$F_{th}$ is the fidelity of teleportation using the thermal state for a cat state of magnitude $|\alpha|=z$, which we have determined numerically. 

\begin{figure*}
\centering
\begin{subfigure}[b]{0.3\textwidth}
    \includegraphics[width=1.15\textwidth]{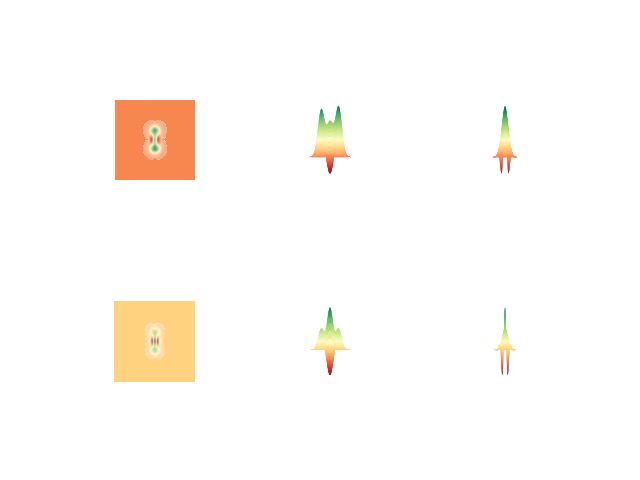}
    \caption{}
    \label{ref:wtap.9}
\end{subfigure}
\hfill
\begin{subfigure}[b]{0.3\textwidth}
    \includegraphics[width=1.15\textwidth]{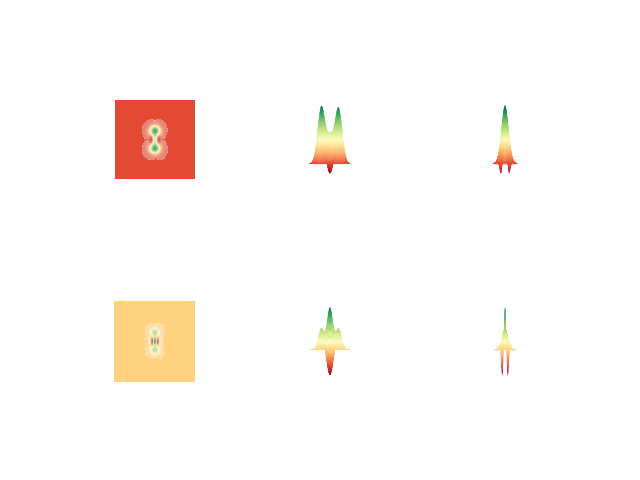}
    \caption{}
    \label{ref:wtap.75}
\end{subfigure}
\hfill
\begin{subfigure}[b]{0.3\textwidth}
     \includegraphics[width=1.15\textwidth]{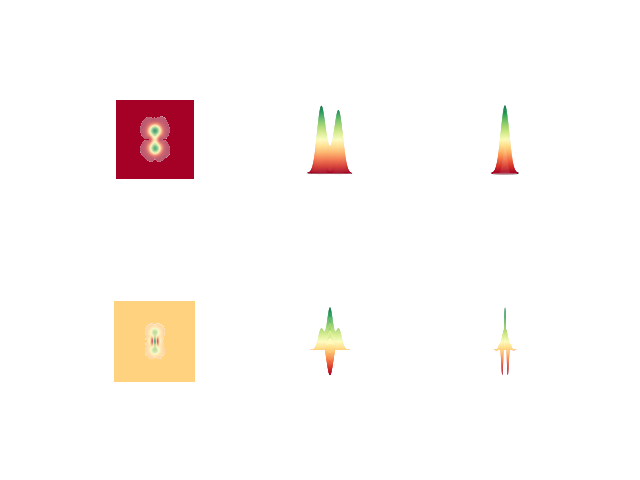}
     \caption{}
     \label{ref:wtap.5}
\end{subfigure}
    
    \caption{Average result of teleportation of the odd $\alpha=-1.5i$ cat state (top row) compared to the original cat state (bottom row), where the TMSV states are being wiretapped (a) by an eavesdropper with transmissivity $\eta=.9$. Average fidelity in this case is about .58 and negativity can clearly be observed in the state. The same follows for (b) $\eta=.75$, with a fidelity of .47, and at (c) $\eta=.5$, with a fidelity of .34, we can no longer observe negativity.  }
    \label{fig:cattelwloss}
\end{figure*}
Another type of state which we can teleport is a GKP state. Such a state encodes a qubit using a lattice, usually a square lattice, distributed in phase space:
\begin{equation}
\begin{aligned}
    \ket{0_L}&\propto \sum_{k=-\infty}^\infty \ket{x=2k\sqrt{\hbar\pi}}\\
    \ket{1_L}&\propto \sum_{k=-\infty}^\infty \ket{x=(2k+1)\sqrt{\hbar\pi}}
\end{aligned}
\end{equation} for the standard lattice spacing length $\sqrt{\pi\hbar}$, where $\ket{x}$ is the eigenstate of the $X$ quadrature operator with eigenvalue $x$. In practice, this state is not physically realizable, so for the purposes of this work we will instead use the realistic GKP state given by $\ket{0}_\epsilon=e^{-\epsilon \hat{n}}\ket{0}_{GKP}$ for $\epsilon=.1$, where $\hat{n}$ is the Fock number operator, except where otherwise noted; this replaces the delta functions above with Gaussians. The GKP state is a good candidate for teleportation because it is naturally robust to the types of displacement errors that occur in the teleportation protocol, and such errors can moreover be actively corrected. Furthermore, if we can teleport, with high qubit fidelity, one mode of an encoded Bell state, we can use this CV channel to transmit qubit entanglement. With that comes the ability to do entanglement concentration or distillation, which are well-understood in the qubit case. 

To justify why Bell state qubits can be teleported in this fashion, we can make the following argument. It is easy to devise a protocol to transmit half of a qubit Bell pair using a second, shared Bell pair as a resource. In the limit of infinite squeezing $r\to\infty$, continuous variable teleportation has the same efficacy as discrete variable teleportation --- it is effectively a maximally-entangled state, since the uncertainty of the quadrature observable is 0 (the same is true of continuous-variable superdense coding). Therefore the only impediment to qubit teleportation is the error resulting from the finite squeezing of the TMSV states, and possibly from loss in the channel. It is shown in \cite{quntao} that these errors form an additive thermal loss channel, and in \cite{noh} that the additive thermal loss channel on a GKP state is correctable. Therefore, we can transmit half of an entangled Bell pair in this way, assuming a sufficiently low effective noise rate as specified in the aforementioned works. 

\begin{figure*}
\centering
    \begin{minipage}[b]{0.45\linewidth}
    \centering
    \includegraphics[trim={0 14cm 3cm 10cm}, scale=.5]{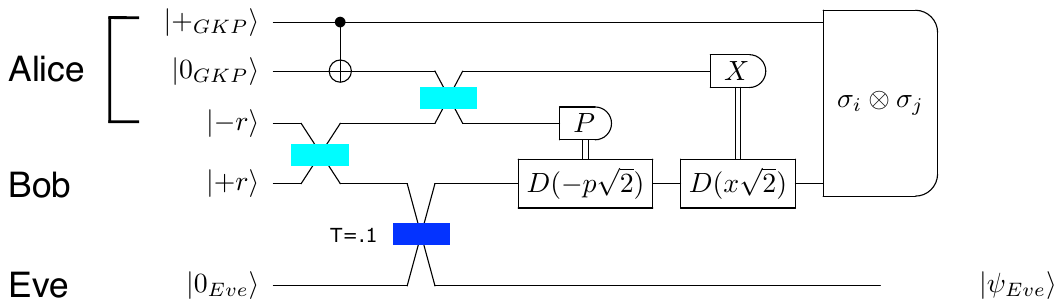}
    \end{minipage}
    \hfill
    \begin{minipage}[b]{0.45\linewidth}
    \centering
    \begin{tabular}{ |p{1cm}||p{1.5cm}|p{1.5cm}|p{1.5cm}|p{1.6cm}|}
 \hline
 \multicolumn{5}{|c|}{Expectations of Paulis} \\
 \hline
 Pauli&Noiseless&Rotated &W+R (.1)&W+R (.01)\\
 \hline
 XX&0.9928&0.6320&0.3238&0.6189\\
 XZ&0.0239&0.6624&0.3396&0.6692\\
 ZX&-0.0147&-0.6002&-0.3100&-0.5744\\
ZZ&0.9893&0.6570&0.3458&0.6557\\
\hline 
S&1.991&2.551&1.319&2.518\\
 \hline
\end{tabular}
    \end{minipage}
    \hfill
    \caption{Using the StrawberryFields package, we were able to simulate the results of two-qubit quantum state tomography on the teleported GKP state using a TMSV state with a high squeezing parameter $r=3.2$. The procedure (with a wiretap of transmissivity .1) is displayed in the above figure; the CNOT in the diagram is a logical CNOT, the $p$ displacement is in the imaginary ($p$) direction, and the measurements at the end are for measuring the Bell inequality violations. The reconstructed matrix was mapped to the nearest density operator in Frobenius norm by taking its positive-semidefinite part, which had a fidelity of .584 with the original two-qubit Bell state and displayed an entanglement of formation of .838. It was verified to be entangled by the Peres-Horodecki criterion \cite{ppt}. It was also possible to measure Bell inequality violations by applying a $3\pi/4$ Pauli Y rotation on the first qubit before the Pauli measurement, either with or without the wiretap, as seen in the table above. With a loss of 0.1, we are not able to observe a Bell inequality violation, but we are able to observe a violation when the loss is instead 0.01. This demonstrates that under ideal conditions, qubit entanglement can be shared steganographically.}
    \label{fig:bellgkptel}
\end{figure*}

\subsubsection{\label{sec:wother} Superdense Coding}
There are some other common tasks which make use of entanglement for purposes of communication, such as superdense coding. It should be noted that continuous variable superdense coding via sending TMSV states through the monitored channel is not a task which can be performed steganographically, even in principle (at least with a perfect fidelity). This is because superdense coding requires Alice to send two correlated instances of information to Bob, and these correlations break the assumption of independence of different uses of the channel which underlies the secrecy guarantees we have previously made. As such it is in principle possible for Eve to determine that the channel is not thermal by doing a joint measurement on two correlated modes.   

\begin{figure}
    \includegraphics[scale=.6]{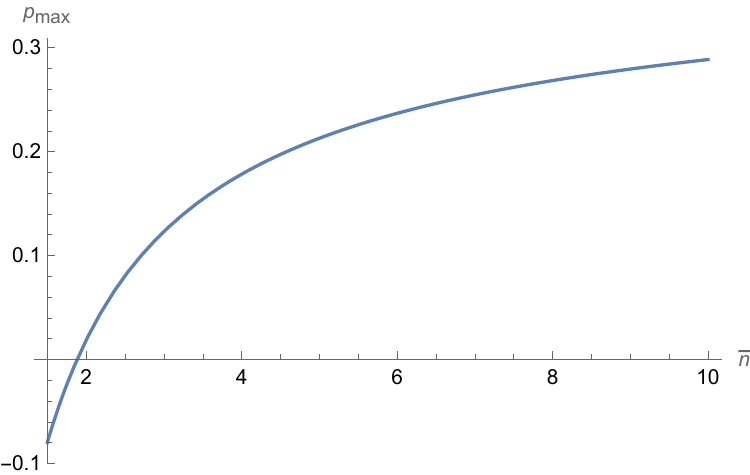}
    \caption{The maximum measurement probability ($p_{\text{max}}$) at which a quantum advantage for superdense coding is still possible, as a function of $\Bar{n}=\sinh^2(r)$ of the mimicked thermal state under the Werner channel.  The value of $p_{\text{max}}$ cannot actually be negative, but that region of the plot indicates that no quantum communication advantage is possible below $r=1.89$.}
    \label{fig:sdensewerner}
\end{figure}

However, if in practice Alice and Bob can establish that they share an entangled TMSV state between them (i.e. Eve is not still holding one of the modes; this likely requires two-way communication, which has not been necessary until now), then a continuous variable superdense coding protocol such as \cite{bksdc} can be performed using two sets of TMSV states. Such a protocol has an asymptotic (in terms of the squeezing parameter) communication rate:
\begin{equation}
    C=\ln(1+\Bar{n}+\Bar{n}^2)\to 4r,
\end{equation}
which is asymptotically twice the classical communication rate. The latter is given by
\begin{equation}
    S(\Bar{n})=(1+\Bar{n})\ln(1+\Bar{n})-\Bar{n}\ln\Bar{n}\to2r.
\end{equation} The condition for outperforming the classical communication rate in this context is thus
\begin{equation}
    (1-p)\ln(1+\Bar{n}+\Bar{n}^2)>(1+\Bar{n})\ln(1+\Bar{n})-\Bar{n}\ln\Bar{n},
\end{equation}
which is satisfiable from about $\Bar{n}>1.89\implies r>1.13$ onwards, as we can see in Figure \ref{fig:sdensewerner}. Below this level of squeezing, continuous variable superdense coding provides no advantage even if $p=0$, and above this squeezing level the maximum acceptable measurement probability $p_{\text{max}}$ to observe a quantum advantage is still lower than for cat state teleportation, likely because Wigner negativity still displays prominently when averaged over, whereas the performance of the channel over each channel use contributes directly and equally to the average rate for superdense coding. Nevertheless, it is a useful measure of quantum communication advantage in this context.

Note that covert superdense coding is only possible because under the Werner eavesdropper model, Eve retains no correlations with the state after she releases it back into the channel, and therefore cannot detect the correlation between the two TMSV modes used for the superdense coding protocol under the above assumptions. Covert superdense coding will not be possible for the wiretap model, as we will show.  
\subsection{\label{sec:wtap}Wiretap Model of the Eavesdropper}
The wiretap model of the eavesdropper is one where the input mode is coupled by a beamsplitter of transmissivity $\eta$ to a mode in the vacuum state. This is equivalent to a pure loss channel of parameter $\eta$; you could say that the model assumes all losses go to Eve. Unlike in the previous model, in this case we can actually correct errors that occur due to the eavesdropper's interference as not all information is destroyed. In fact, GKP and cat codes are both known to be able to correct (at least approximately) errors resulting from the pure loss channel. In this case it is the TMSV state that is experiencing loss, not the encoded state itself; nevertheless, the effective channel acting on the GKP state is still Gaussian and therefore correctable. However, Alice and Bob must be careful what channel operations they perform on the states, as in principle Eve could be storing the tapped portions of the states to use in a joint measurement later. Without knowing the true loss in the channel, Alice and Bob are unsure whether Eve is tapping the channel and how much, which is similar to the underlying uncertainty of the true channel state that makes steganography possible (only in this case Eve is the one who is hidden).
\subsubsection{State Teleportation}
If we reexamine the cat state teleportation protocol, we see in Figure \ref{fig:cattelwloss} that at $\eta=.5$ there is no longer negativity to be observed. This is intuitive: you can imagine that Bob no longer has a communication advantage against Eve, and therefore there cannot be a quantum advantage either. 

To formalize this argument, we can examine the notion of degradable and anti-degradable channels \cite{shordev,antidegrade}. A degradable channel between Alice, Bob, and Eve $\mathcal{N}_{AB}$ is one where there exists a completely positive trace-preserving (CPTP) map $\Phi$ that when applied to the channel $\mathcal{N}_{AB}$ gives the complementary channel $\mathcal{N}_{AE}=\Phi(\mathcal{N}_{AB})$. In this case, the complementary channel $\mathcal{N}_{AE}$ is called anti-degradable. The wiretap channel with $\eta>.5$, the results of which we see in Figure \ref{fig:cattelwloss}, is a degradable channel \cite{Wilde}. This can be easily seen by passing the result of the degradable channel with $\eta>.5$ through another beam-splitter of transmissivity $\eta'$ so that the $(\eta)(\eta')=1-\eta$. If the wiretap channel with $\mathcal{N}_{AB}(\eta)$ is degradable, the wiretap channel with $\mathcal{N}_{AB}(\eta')$ is anti-degradable for $\eta'<.5$. Therefore, its quantum capacity is 0 (otherwise we could violate the no-cloning theorem by having both the channel and the complementary channel teleport a given quantum state). Since this is the case, it is impossible to observe any signature of quantum advantage in the teleported state, including negativity. 

\begin{figure}
    \includegraphics[scale=.35]{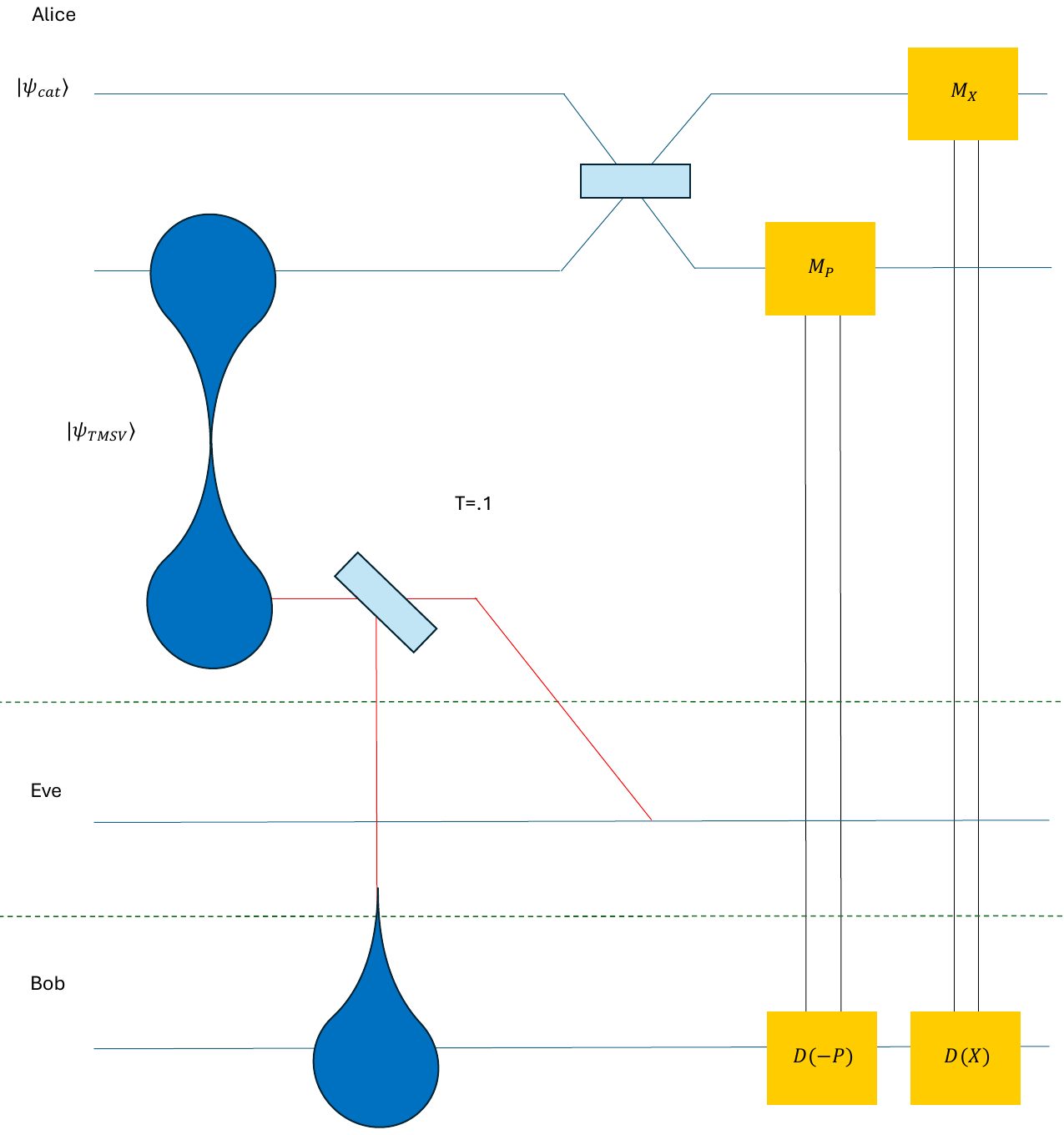}
    \includegraphics[scale=.25]{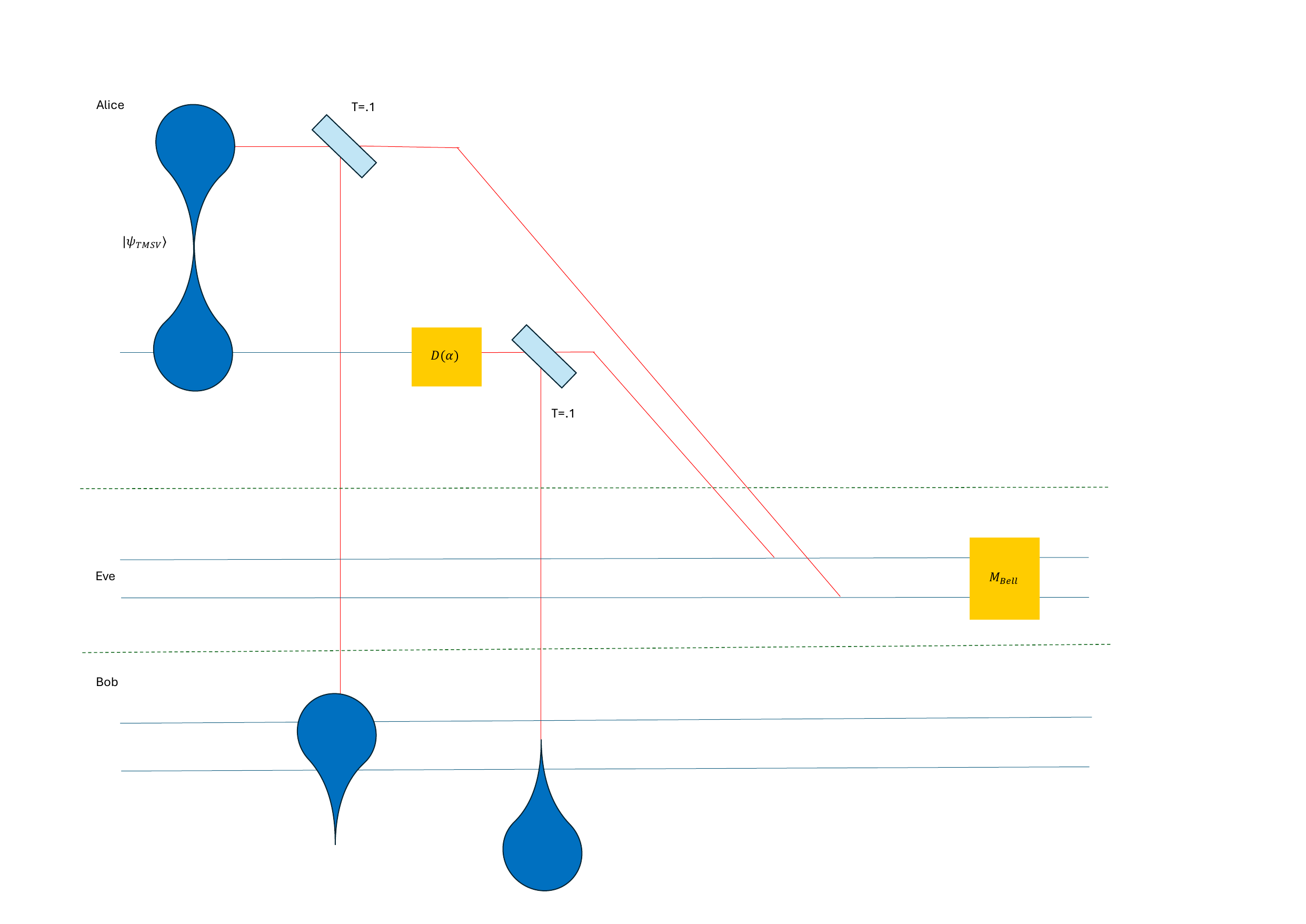}
    \caption{The circuits for continuous variable teleportation (top) and superdense coding (bottom) under the wiretap model. In the former case, Eve can only interfere with Alice and Bob's operation of the circuit, but in the latter case, since two correlated bits of information pass through the channel, Eve can in principle detect quantum correlations via a Bell-type or other joint measurement.}
    \label{fig:wtapeve}
\end{figure}
By referencing the Pirandola-Laurenza-Ottaviani-Banchi(PLOB) bound \cite{plob}, we can expect that with a transmissivity of $.9$ and therefore an achievable rate of $-\log_2(.1)=3.32$ it should be possible to transmit qubit entanglement via teleportation. This can be done using GKP states, for example. In Fig. \ref{fig:bellgkptel} we describe the results of simulating the teleportation of GKP states using a TMSV state as the resource. By reconstructing the density matrix as described there, we are able to verify whether the state is entangled by using the Peres-Horodecki criterion. We further verify the presence of entanglement by measuring Bell inequality violations on a rotated version of the teleported GKP bell state; the rotation is a Pauli Y rotation by $3\pi/4$, which maximizes the violation of the Bell inequality with the standard expectations of $XX$, $XZ$, $ZX$, and $ZZ$ operators. Indeed, we are able to observe a violation with $S=2.55$, which is roughly in line with the concurrence of the reconstructed state which was calculated to be .89. If the wiretap is in place, the state decoheres significantly and a Bell inequality violation is not observable. It may be possible to transmit qubit entanglement through the wiretap with $\eta=.1$ using a distillation or concentration procedure, but the above demonstration of qubit entanglement sharing demonstrates that for a sufficiently low level of wiretapping, steganographic qubit entanglement sharing is possible.   
\subsubsection{Superdense Coding}
Figure \ref{fig:wtapeve} shows that steganographic superdense coding is impossible under this eavesdropper model, since
you must always send correlated information through the
channel which Eve can store and detect using a joint
measurement. This is in contrast to the Werner channel;
while in theory Eve could hold a state in that channel as
well, Alice and Bob would be able to detect that the state
they have is uncorrelated, whereas holding sufficient information to detect the sending of correlated information in
the wiretap scheme does not produce the same signature
under wiretap assumptions.

It is possible for Alice and Bob to circumvent this
issue, however, if they can communicate classically between them (both ways). A sketch of the procedure for
doing so would be as follows: first, Alice sends one mode
of the TMSV state through the channel. Eve siphons
off part of this state using her beam splitter, while Bob
receives the remainder. Then, Alice does not send the
other mode of the TMSV state through the channel, but
instead repeats this procedure with another TMSV state,
however many times will be required for eventual distillation. Eve now has several uncorrelated fragments of
TMSV states. Alice and Bob then communicate covertly,
perhaps using classical steganography, so that they can
distill the states into a pure TMSV state, using a procedure such as what is described in \cite{distill}. In so doing, they
eliminate the correlations between the original mode that
Eve siphoned, and the remaining modes Alice has available to send. Alice and Bob can now use the other mode of each TMSV state
for superdense coding normally. In this series of events,
the degree of wiretapping dictates the success probability
for the distillation procedure.

\section{Conclusion}
In this work, we have shown that steganographic entanglement sharing is possible using standard TMSV states under two eavesdropper models. We have also verified that even without entanglement concentration, this capability is sufficient for tasks such as teleportation and qubit teleportation at a level that still demonstrates a quantum advantage despite the presence of an eavesdropper. There is a limitation to these protocols that would be interesting for future study: TMSV state generation is possible using a periodically poled crystal or single mode squeezing, but this limits the wavelength of the states to the relevant wavelengths for which these processes are possible. Put another way: it is only possible to mimic thermal states with certain convenient values of $\Bar{n}$. This is unlike the coherent state steganography studied previously in \cite{priorwork}, where in principle it was possible to emulate a thermal state of any $\Bar{n}$ using modulation as was shown in \cite{priorexperiment}. Thus, to enhance the secrecy it may be interesting to consider methods of up- and down- conversion, to more readily match thermal states emitted across a variety of possible wavelengths \cite{upconv}. The discussion of the effects of such schemes on secrecy or communication capacity are left for a future work. 
\begin{acknowledgments}
The authors wish to acknowledge the support of NSF Grants 1719778, 1911089 and 2316713. BA would also like to thank David Wei, Sidarth Ramanujan Raghunathan, and Anirudh Lanka for helpful discussions.
\end{acknowledgments}
\appendix

\section{Fuchs-Van de Graaf Lower Bound for a Mixture of Two States}
\label{appendixa}

Consider a density operator $\rho_0$, and a second density operator $\rho_1=p\rho_{th}+(1-p)\rho_c$ for two valid density operators $\rho_{th}$ and $\rho_c$. A direct application of the Fuchs-Van de Graaf Bound \cite{FVDGbound} gives
\begin{equation}
\begin{aligned}
    1-\sqrt{F(\rho_1, \rho_0)}\leq& \frac{1}{2}||\rho_1-\rho_0||\\
      F(\rho_1, \rho_0)\geq&\left(1-\frac{1}{2}||\rho_1-\rho_0||\right)^2\\
\end{aligned}
\end{equation}
to which we can apply the triangle inequality as follows: 
\begin{equation}
    \begin{aligned}
      F(\rho_1, \rho_0)\geq&\left(1-\frac{1}{2}||p(\rho_{th}-\rho_0)+(1-p)(\rho_c-\rho_0)||\right)^2\\
      F(\rho_1, \rho_0)\geq&\left(1-\frac{p}{2}||\rho_{th}-\rho_0||-\frac{1-p}{2}||\rho_c-\rho_0||\right)^2\\
      F(\rho_1, \rho_0)\geq&\Bigg(p\left(1-\frac{1}{2}||\rho_{th}-\rho_0||\right)\\&+(1-p)\left(1-\frac{1}{2}||\rho_c-\rho_0||\right)\Bigg)^2,
\end{aligned}
\end{equation} where the above norm is the trace norm. Therefore the lower bounds on the individual states of the mixture can be used to define a lower bound on the mixture, which greatly increases computational efficiency. Note that a similar derivation is not possible for the upper bound, but that the upper bound does hold whenever $\rho_1$ is pure (for example, if $\rho_c$ is a pure state it holds for $p=0$), and therefore since the tightness of the upper bound decreases as the purity of the states decrease (by contrast with the lower bound), it may still hold in practice or be useful for illustrative purposes.


\bibliography{bibliography}

\providecommand{\noopsort}[1]{}\providecommand{\singleletter}[1]{#1}%
\begin{thebibliography}{23}%
\makeatletter
\providecommand \@ifxundefined [1]{%
 \@ifx{#1\undefined}
}%
\providecommand \@ifnum [1]{%
 \ifnum #1\expandafter \@firstoftwo
 \else \expandafter \@secondoftwo
 \fi
}%
\providecommand \@ifx [1]{%
 \ifx #1\expandafter \@firstoftwo
 \else \expandafter \@secondoftwo
 \fi
}%
\providecommand \natexlab [1]{#1}%
\providecommand \enquote  [1]{``#1''}%
\providecommand \bibnamefont  [1]{#1}%
\providecommand \bibfnamefont [1]{#1}%
\providecommand \citenamefont [1]{#1}%
\providecommand \href@noop [0]{\@secondoftwo}%
\providecommand \href [0]{\begingroup \@sanitize@url \@href}%
\providecommand \@href[1]{\@@startlink{#1}\@@href}%
\providecommand \@@href[1]{\endgroup#1\@@endlink}%
\providecommand \@sanitize@url [0]{\catcode `\\12\catcode `\$12\catcode
  `\&12\catcode `\#12\catcode `\^12\catcode `\_12\catcode `\%12\relax}%
\providecommand \@@startlink[1]{}%
\providecommand \@@endlink[0]{}%
\providecommand \url  [0]{\begingroup\@sanitize@url \@url }%
\providecommand \@url [1]{\endgroup\@href {#1}{\urlprefix }}%
\providecommand \urlprefix  [0]{URL }%
\providecommand \Eprint [0]{\href }%
\providecommand \doibase [0]{https://doi.org/}%
\providecommand \selectlanguage [0]{\@gobble}%
\providecommand \bibinfo  [0]{\@secondoftwo}%
\providecommand \bibfield  [0]{\@secondoftwo}%
\providecommand \translation [1]{[#1]}%
\providecommand \BibitemOpen [0]{}%
\providecommand \bibitemStop [0]{}%
\providecommand \bibitemNoStop [0]{.\EOS\space}%
\providecommand \EOS [0]{\spacefactor3000\relax}%
\providecommand \BibitemShut  [1]{\csname bibitem#1\endcsname}%
\let\auto@bib@innerbib\@empty
\bibitem [{\citenamefont {Avritzer}\ and\ \citenamefont
  {Brun}(2024)}]{priorwork}%
  \BibitemOpen
  \bibfield  {author} {\bibinfo {author} {\bibfnamefont {B.}~\bibnamefont
  {Avritzer}}\ and\ \bibinfo {author} {\bibfnamefont {T.~A.}\ \bibnamefont
  {Brun}},\ }\bibfield  {title} {\bibinfo {title} {Quantum steganography via
  coherent- and fock-state encoding in an optical medium},\ }\href
  {https://doi.org/10.1103/PhysRevA.109.032401} {\bibfield  {journal} {\bibinfo
   {journal} {Phys. Rev. A}\ }\textbf {\bibinfo {volume} {109}},\ \bibinfo
  {pages} {032401} (\bibinfo {year} {2024})}\BibitemShut {NoStop}%
\bibitem [{\citenamefont {Wu}\ and\ \citenamefont {Narimanov}(2006)}]{Wu1}%
  \BibitemOpen
  \bibfield  {author} {\bibinfo {author} {\bibfnamefont {B.~B.}\ \bibnamefont
  {Wu}}\ and\ \bibinfo {author} {\bibfnamefont {E.~E.}\ \bibnamefont
  {Narimanov}},\ }\bibfield  {title} {\bibinfo {title} {A method for secure
  communications over a public fiber-optical network},\ }\href
  {https://doi.org/10.1364/OE.14.003738} {\bibfield  {journal} {\bibinfo
  {journal} {Opt. Express}\ }\textbf {\bibinfo {volume} {14}},\ \bibinfo
  {pages} {3738} (\bibinfo {year} {2006})}\BibitemShut {NoStop}%
\bibitem [{\citenamefont {Wu}\ \emph {et~al.}(2013)\citenamefont {Wu},
  \citenamefont {Wang}, \citenamefont {Tian}, \citenamefont {Fok},
  \citenamefont {Shastri}, \citenamefont {Kanoff},\ and\ \citenamefont
  {Prucnal}}]{Wu2}%
  \BibitemOpen
  \bibfield  {author} {\bibinfo {author} {\bibfnamefont {B.}~\bibnamefont
  {Wu}}, \bibinfo {author} {\bibfnamefont {Z.}~\bibnamefont {Wang}}, \bibinfo
  {author} {\bibfnamefont {Y.}~\bibnamefont {Tian}}, \bibinfo {author}
  {\bibfnamefont {M.~P.}\ \bibnamefont {Fok}}, \bibinfo {author} {\bibfnamefont
  {B.~J.}\ \bibnamefont {Shastri}}, \bibinfo {author} {\bibfnamefont {D.~R.}\
  \bibnamefont {Kanoff}},\ and\ \bibinfo {author} {\bibfnamefont {P.~R.}\
  \bibnamefont {Prucnal}},\ }\bibfield  {title} {\bibinfo {title} {Optical
  steganography based on amplified spontaneous emission noise},\ }\href
  {https://doi.org/10.1364/OE.21.002065} {\bibfield  {journal} {\bibinfo
  {journal} {Opt. Express}\ }\textbf {\bibinfo {volume} {21}},\ \bibinfo
  {pages} {2065} (\bibinfo {year} {2013})}\BibitemShut {NoStop}%
\bibitem [{\citenamefont {Dutt}\ \emph {et~al.}(2015)\citenamefont {Dutt},
  \citenamefont {Luke}, \citenamefont {Manipatruni}, \citenamefont {Gaeta},
  \citenamefont {Nussenzveig},\ and\ \citenamefont {Lipson}}]{sqexp}%
  \BibitemOpen
  \bibfield  {author} {\bibinfo {author} {\bibfnamefont {A.}~\bibnamefont
  {Dutt}}, \bibinfo {author} {\bibfnamefont {K.}~\bibnamefont {Luke}}, \bibinfo
  {author} {\bibfnamefont {S.}~\bibnamefont {Manipatruni}}, \bibinfo {author}
  {\bibfnamefont {A.~L.}\ \bibnamefont {Gaeta}}, \bibinfo {author}
  {\bibfnamefont {P.}~\bibnamefont {Nussenzveig}},\ and\ \bibinfo {author}
  {\bibfnamefont {M.}~\bibnamefont {Lipson}},\ }\bibfield  {title} {\bibinfo
  {title} {On-chip optical squeezing},\ }\href@noop {} {\bibfield  {journal}
  {\bibinfo  {journal} {Phys. Rev. Applied}\ }\textbf {\bibinfo {volume} {3}},\
  \bibinfo {pages} {044005} (\bibinfo {year} {2015})}\BibitemShut {NoStop}%
\bibitem [{\citenamefont {Fuchs}\ and\ \citenamefont {van~de
  Graaf}(1997)}]{FVDGbound}%
  \BibitemOpen
  \bibfield  {author} {\bibinfo {author} {\bibfnamefont {C.~A.}\ \bibnamefont
  {Fuchs}}\ and\ \bibinfo {author} {\bibfnamefont {J.}~\bibnamefont {van~de
  Graaf}},\ }\bibfield  {title} {\bibinfo {title} {Cryptographic
  distinguishability measures for quantum mechanical states},\ }\href@noop {}
  {\bibfield  {journal} {\bibinfo  {journal} {""}\ } (\bibinfo {year}
  {1997})},\ \Eprint {https://arxiv.org/abs/arXiv:quant-ph/9712042}
  {arXiv:quant-ph/9712042} \BibitemShut {NoStop}%
\bibitem [{\citenamefont {Killoran}\ \emph {et~al.}(2019)\citenamefont
  {Killoran}, \citenamefont {Izaac}, \citenamefont {Quesada}, \citenamefont
  {Bergholm}, \citenamefont {Amy},\ and\ \citenamefont {Weedbrook}}]{sfields1}%
  \BibitemOpen
  \bibfield  {author} {\bibinfo {author} {\bibfnamefont {N.}~\bibnamefont
  {Killoran}}, \bibinfo {author} {\bibfnamefont {J.}~\bibnamefont {Izaac}},
  \bibinfo {author} {\bibfnamefont {N.}~\bibnamefont {Quesada}}, \bibinfo
  {author} {\bibfnamefont {V.}~\bibnamefont {Bergholm}}, \bibinfo {author}
  {\bibfnamefont {M.}~\bibnamefont {Amy}},\ and\ \bibinfo {author}
  {\bibfnamefont {C.}~\bibnamefont {Weedbrook}},\ }\bibfield  {title} {\bibinfo
  {title} {Strawberry {F}ields: {A} {S}oftware {P}latform for {P}hotonic
  {Q}uantum {C}omputing},\ }\href {https://doi.org/10.22331/q-2019-03-11-129}
  {\bibfield  {journal} {\bibinfo  {journal} {{Quantum}}\ }\textbf {\bibinfo
  {volume} {3}},\ \bibinfo {pages} {129} (\bibinfo {year} {2019})}\BibitemShut
  {NoStop}%
\bibitem [{\citenamefont {Bromley}\ \emph {et~al.}(2020)\citenamefont
  {Bromley}, \citenamefont {Arrazola}, \citenamefont {Jahangiri}, \citenamefont
  {Izaac}, \citenamefont {Quesada}, \citenamefont {Gran}, \citenamefont
  {Schuld}, \citenamefont {Swinarton}, \citenamefont {Zabaneh},\ and\
  \citenamefont {Killoran}}]{sfields2}%
  \BibitemOpen
  \bibfield  {author} {\bibinfo {author} {\bibfnamefont {T.~R.}\ \bibnamefont
  {Bromley}}, \bibinfo {author} {\bibfnamefont {J.~M.}\ \bibnamefont
  {Arrazola}}, \bibinfo {author} {\bibfnamefont {S.}~\bibnamefont {Jahangiri}},
  \bibinfo {author} {\bibfnamefont {J.}~\bibnamefont {Izaac}}, \bibinfo
  {author} {\bibfnamefont {N.}~\bibnamefont {Quesada}}, \bibinfo {author}
  {\bibfnamefont {A.~D.}\ \bibnamefont {Gran}}, \bibinfo {author}
  {\bibfnamefont {M.}~\bibnamefont {Schuld}}, \bibinfo {author} {\bibfnamefont
  {J.}~\bibnamefont {Swinarton}}, \bibinfo {author} {\bibfnamefont
  {Z.}~\bibnamefont {Zabaneh}},\ and\ \bibinfo {author} {\bibfnamefont
  {N.}~\bibnamefont {Killoran}},\ }\bibfield  {title} {\bibinfo {title}
  {Applications of near-term photonic quantum computers: software and
  algorithms},\ }\href {https://doi.org/10.1088/2058-9565/ab8504} {\bibfield
  {journal} {\bibinfo  {journal} {Quantum Science and Technology}\ }\textbf
  {\bibinfo {volume} {5}},\ \bibinfo {pages} {034010} (\bibinfo {year}
  {2020})}\BibitemShut {NoStop}%
\bibitem [{\citenamefont {Werner}(1989)}]{Werner}%
  \BibitemOpen
  \bibfield  {author} {\bibinfo {author} {\bibfnamefont {R.~F.}\ \bibnamefont
  {Werner}},\ }\bibfield  {title} {\bibinfo {title} {Quantum states with
  einstein-podolsky-rosen correlations admitting a hidden-variable model},\
  }\href {https://doi.org/10.1103/PhysRevA.40.4277} {\bibfield  {journal}
  {\bibinfo  {journal} {Phys. Rev. A}\ }\textbf {\bibinfo {volume} {40}},\
  \bibinfo {pages} {4277} (\bibinfo {year} {1989})}\BibitemShut {NoStop}%
\bibitem [{\citenamefont {Mi\ifmmode~\check{s}\else \v{s}\fi{}ta}\ \emph
  {et~al.}(2002)\citenamefont {Mi\ifmmode~\check{s}\else \v{s}\fi{}ta},
  \citenamefont {Filip},\ and\ \citenamefont {Fiur\'a\ifmmode~\check{s}\else
  \v{s}\fi{}ek}}]{FWerner}%
  \BibitemOpen
  \bibfield  {author} {\bibinfo {author} {\bibfnamefont {L.}~\bibnamefont
  {Mi\ifmmode~\check{s}\else \v{s}\fi{}ta}}, \bibinfo {author} {\bibfnamefont
  {R.}~\bibnamefont {Filip}},\ and\ \bibinfo {author} {\bibfnamefont
  {J.}~\bibnamefont {Fiur\'a\ifmmode~\check{s}\else \v{s}\fi{}ek}},\ }\bibfield
   {title} {\bibinfo {title} {Continuous-variable werner state: Separability,
  nonlocality, squeezing, and teleportation},\ }\href
  {https://doi.org/10.1103/PhysRevA.65.062315} {\bibfield  {journal} {\bibinfo
  {journal} {Phys. Rev. A}\ }\textbf {\bibinfo {volume} {65}},\ \bibinfo
  {pages} {062315} (\bibinfo {year} {2002})}\BibitemShut {NoStop}%
\bibitem [{\citenamefont {Hastrup}\ and\ \citenamefont
  {Andersen}(2022{\natexlab{a}})}]{eccat}%
  \BibitemOpen
  \bibfield  {author} {\bibinfo {author} {\bibfnamefont {J.}~\bibnamefont
  {Hastrup}}\ and\ \bibinfo {author} {\bibfnamefont {U.~L.}\ \bibnamefont
  {Andersen}},\ }\bibfield  {title} {\bibinfo {title} {All-optical cat-code
  quantum error correction},\ }\href
  {https://doi.org/10.1103/PhysRevResearch.4.043065} {\bibfield  {journal}
  {\bibinfo  {journal} {Phys. Rev. Res.}\ }\textbf {\bibinfo {volume} {4}},\
  \bibinfo {pages} {043065} (\bibinfo {year} {2022}{\natexlab{a}})}\BibitemShut
  {NoStop}%
\bibitem [{\citenamefont {Hastrup}\ and\ \citenamefont
  {Andersen}(2022{\natexlab{b}})}]{catgkp}%
  \BibitemOpen
  \bibfield  {author} {\bibinfo {author} {\bibfnamefont {J.}~\bibnamefont
  {Hastrup}}\ and\ \bibinfo {author} {\bibfnamefont {U.~L.}\ \bibnamefont
  {Andersen}},\ }\bibfield  {title} {\bibinfo {title} {Protocol for generating
  optical gottesman-kitaev-preskill states with cavity qed},\ }\href
  {https://doi.org/10.1103/PhysRevLett.128.170503} {\bibfield  {journal}
  {\bibinfo  {journal} {Phys. Rev. Lett.}\ }\textbf {\bibinfo {volume} {128}},\
  \bibinfo {pages} {170503} (\bibinfo {year} {2022}{\natexlab{b}})}\BibitemShut
  {NoStop}%
\bibitem [{\citenamefont {Braunstein}\ and\ \citenamefont
  {Kimble}(1998)}]{BKTel}%
  \BibitemOpen
  \bibfield  {author} {\bibinfo {author} {\bibfnamefont {S.~L.}\ \bibnamefont
  {Braunstein}}\ and\ \bibinfo {author} {\bibfnamefont {H.~J.}\ \bibnamefont
  {Kimble}},\ }\bibfield  {title} {\bibinfo {title} {Teleportation of
  continuous quantum variables},\ }\href
  {https://doi.org/10.1103/PhysRevLett.80.869} {\bibfield  {journal} {\bibinfo
  {journal} {Phys. Rev. Lett.}\ }\textbf {\bibinfo {volume} {80}},\ \bibinfo
  {pages} {869} (\bibinfo {year} {1998})}\BibitemShut {NoStop}%
\bibitem [{\citenamefont {Wu}\ \emph {et~al.}(2022)\citenamefont {Wu},
  \citenamefont {Zhang},\ and\ \citenamefont {Zhuang}}]{quntao}%
  \BibitemOpen
  \bibfield  {author} {\bibinfo {author} {\bibfnamefont {B.-H.}\ \bibnamefont
  {Wu}}, \bibinfo {author} {\bibfnamefont {Z.}~\bibnamefont {Zhang}},\ and\
  \bibinfo {author} {\bibfnamefont {Q.}~\bibnamefont {Zhuang}},\ }\bibfield
  {title} {\bibinfo {title} {Continuous-variable quantum repeaters based on
  bosonic error-correction and teleportation: architecture and applications},\
  }\href {https://doi.org/10.1088/2058-9565/ac4f6b} {\bibfield  {journal}
  {\bibinfo  {journal} {Quantum Science and Technology}\ }\textbf {\bibinfo
  {volume} {7}},\ \bibinfo {pages} {025018} (\bibinfo {year}
  {2022})}\BibitemShut {NoStop}%
\bibitem [{\citenamefont {Noh}\ \emph {et~al.}(2020)\citenamefont {Noh},
  \citenamefont {Girvin},\ and\ \citenamefont {Jiang}}]{noh}%
  \BibitemOpen
  \bibfield  {author} {\bibinfo {author} {\bibfnamefont {K.}~\bibnamefont
  {Noh}}, \bibinfo {author} {\bibfnamefont {S.~M.}\ \bibnamefont {Girvin}},\
  and\ \bibinfo {author} {\bibfnamefont {L.}~\bibnamefont {Jiang}},\ }\bibfield
   {title} {\bibinfo {title} {Encoding an oscillator into many oscillators},\
  }\href {https://doi.org/10.1103/PhysRevLett.125.080503} {\bibfield  {journal}
  {\bibinfo  {journal} {Phys. Rev. Lett.}\ }\textbf {\bibinfo {volume} {125}},\
  \bibinfo {pages} {080503} (\bibinfo {year} {2020})}\BibitemShut {NoStop}%
\bibitem [{\citenamefont {Horodecki}\ \emph {et~al.}(1996)\citenamefont
  {Horodecki}, \citenamefont {Horodecki},\ and\ \citenamefont
  {Horodecki}}]{ppt}%
  \BibitemOpen
  \bibfield  {author} {\bibinfo {author} {\bibfnamefont {M.}~\bibnamefont
  {Horodecki}}, \bibinfo {author} {\bibfnamefont {P.}~\bibnamefont
  {Horodecki}},\ and\ \bibinfo {author} {\bibfnamefont {R.}~\bibnamefont
  {Horodecki}},\ }\bibfield  {title} {\bibinfo {title} {Separability of mixed
  states: necessary and sufficient conditions},\ }\href
  {https://doi.org/https://doi.org/10.1016/S0375-9601(96)00706-2} {\bibfield
  {journal} {\bibinfo  {journal} {Physics Letters A}\ }\textbf {\bibinfo
  {volume} {223}},\ \bibinfo {pages} {1} (\bibinfo {year} {1996})}\BibitemShut
  {NoStop}%
\bibitem [{\citenamefont {Braunstein}\ and\ \citenamefont
  {Kimble}(2000)}]{bksdc}%
  \BibitemOpen
  \bibfield  {author} {\bibinfo {author} {\bibfnamefont {S.~L.}\ \bibnamefont
  {Braunstein}}\ and\ \bibinfo {author} {\bibfnamefont {H.~J.}\ \bibnamefont
  {Kimble}},\ }\bibfield  {title} {\bibinfo {title} {Dense coding for
  continuous variables},\ }\href {https://doi.org/10.1103/PhysRevA.61.042302}
  {\bibfield  {journal} {\bibinfo  {journal} {Phys. Rev. A}\ }\textbf {\bibinfo
  {volume} {61}},\ \bibinfo {pages} {042302} (\bibinfo {year}
  {2000})}\BibitemShut {NoStop}%
\bibitem [{\citenamefont {Devetak}\ and\ \citenamefont {Shor}(2005)}]{shordev}%
  \BibitemOpen
  \bibfield  {author} {\bibinfo {author} {\bibfnamefont {I.}~\bibnamefont
  {Devetak}}\ and\ \bibinfo {author} {\bibfnamefont {P.~W.}\ \bibnamefont
  {Shor}},\ }\bibfield  {title} {\bibinfo {title} {The capacity of a quantum
  channel for simultaneous transmission of classical and quantum information},\
  }\href@noop {} {\bibfield  {journal} {\bibinfo  {journal} {Commun. Math.
  Phys.}\ }\textbf {\bibinfo {volume} {256}},\ \bibinfo {pages} {287} (\bibinfo
  {year} {2005})}\BibitemShut {NoStop}%
\bibitem [{\citenamefont {Cubitt}\ \emph {et~al.}(2008)\citenamefont {Cubitt},
  \citenamefont {Ruskai},\ and\ \citenamefont {Smith}}]{antidegrade}%
  \BibitemOpen
  \bibfield  {author} {\bibinfo {author} {\bibfnamefont {T.~S.}\ \bibnamefont
  {Cubitt}}, \bibinfo {author} {\bibfnamefont {M.~B.}\ \bibnamefont {Ruskai}},\
  and\ \bibinfo {author} {\bibfnamefont {G.}~\bibnamefont {Smith}},\ }\bibfield
   {title} {\bibinfo {title} {The structure of degradable quantum channels},\
  }\href@noop {} {\bibfield  {journal} {\bibinfo  {journal} {J. Math. Phys.}\
  }\textbf {\bibinfo {volume} {49}},\ \bibinfo {pages} {102104} (\bibinfo
  {year} {2008})}\BibitemShut {NoStop}%
\bibitem [{\citenamefont {Wilde}(2013)}]{Wilde}%
  \BibitemOpen
  \bibfield  {author} {\bibinfo {author} {\bibfnamefont {M.~M.}\ \bibnamefont
  {Wilde}},\ }\href@noop {} {\emph {\bibinfo {title} {Quantum Information
  Theory}}}\ (\bibinfo  {publisher} {Cambridge University Press},\ \bibinfo
  {year} {2013})\BibitemShut {NoStop}%
\bibitem [{\citenamefont {Pirandola}\ \emph {et~al.}(2017)\citenamefont
  {Pirandola}, \citenamefont {Laurenza}, \citenamefont {Ottaviani},\ and\
  \citenamefont {Banchi}}]{plob}%
  \BibitemOpen
  \bibfield  {author} {\bibinfo {author} {\bibfnamefont {S.}~\bibnamefont
  {Pirandola}}, \bibinfo {author} {\bibfnamefont {R.}~\bibnamefont {Laurenza}},
  \bibinfo {author} {\bibfnamefont {C.}~\bibnamefont {Ottaviani}},\ and\
  \bibinfo {author} {\bibfnamefont {L.}~\bibnamefont {Banchi}},\ }\bibfield
  {title} {\bibinfo {title} {Fundamental limits of repeaterless quantum
  communications},\ }\href@noop {} {\bibfield  {journal} {\bibinfo  {journal}
  {Nat. Commun.}\ }\textbf {\bibinfo {volume} {8}},\ \bibinfo {pages} {15043}
  (\bibinfo {year} {2017})}\BibitemShut {NoStop}%
\bibitem [{\citenamefont {Seshadreesan}\ \emph {et~al.}(2019)\citenamefont
  {Seshadreesan}, \citenamefont {Krovi},\ and\ \citenamefont {Guha}}]{distill}%
  \BibitemOpen
  \bibfield  {author} {\bibinfo {author} {\bibfnamefont {K.~P.}\ \bibnamefont
  {Seshadreesan}}, \bibinfo {author} {\bibfnamefont {H.}~\bibnamefont
  {Krovi}},\ and\ \bibinfo {author} {\bibfnamefont {S.}~\bibnamefont {Guha}},\
  }\bibfield  {title} {\bibinfo {title} {Continuous-variable entanglement
  distillation over a pure loss channel with multiple quantum scissors},\
  }\bibfield  {journal} {\bibinfo  {journal} {Physical Review A}\ }\textbf
  {\bibinfo {volume} {100}},\ \href
  {https://doi.org/10.1103/physreva.100.022315} {10.1103/physreva.100.022315}
  (\bibinfo {year} {2019})\BibitemShut {NoStop}%
\bibitem [{\citenamefont {Weinstein}\ \emph {et~al.}(2024)\citenamefont
  {Weinstein}, \citenamefont {Avritzer}, \citenamefont {Brun},\ and\
  \citenamefont {Habif}}]{priorexperiment}%
  \BibitemOpen
  \bibfield  {author} {\bibinfo {author} {\bibfnamefont {H.}~\bibnamefont
  {Weinstein}}, \bibinfo {author} {\bibfnamefont {B.}~\bibnamefont {Avritzer}},
  \bibinfo {author} {\bibfnamefont {T.~A.}\ \bibnamefont {Brun}},\ and\
  \bibinfo {author} {\bibfnamefont {J.~L.}\ \bibnamefont {Habif}},\ }\href@noop
  {} {\bibinfo {title} {High fidelity artificial quantum thermal state
  generation using encoded coherent states}} (\bibinfo {year} {2024}),\ \Eprint
  {https://arxiv.org/abs/2405.03881} {arXiv:2405.03881 [quant-ph]} \BibitemShut
  {NoStop}%
\bibitem [{\citenamefont {McKinstrie}\ \emph {et~al.}(2005)\citenamefont
  {McKinstrie}, \citenamefont {Harvey}, \citenamefont {Radic},\ and\
  \citenamefont {Raymer}}]{upconv}%
  \BibitemOpen
  \bibfield  {author} {\bibinfo {author} {\bibfnamefont {C.~J.}\ \bibnamefont
  {McKinstrie}}, \bibinfo {author} {\bibfnamefont {J.~D.}\ \bibnamefont
  {Harvey}}, \bibinfo {author} {\bibfnamefont {S.}~\bibnamefont {Radic}},\ and\
  \bibinfo {author} {\bibfnamefont {M.~G.}\ \bibnamefont {Raymer}},\ }\bibfield
   {title} {\bibinfo {title} {Translation of quantum states by four-wave mixing
  in fibers},\ }\href {https://doi.org/10.1364/OPEX.13.009131} {\bibfield
  {journal} {\bibinfo  {journal} {Opt. Express}\ }\textbf {\bibinfo {volume}
  {13}},\ \bibinfo {pages} {9131} (\bibinfo {year} {2005})}\BibitemShut
  {NoStop}%
\end{thebibliography}%

\end{document}